# Inter-cluster reactivity of Metallo-aromatic and anti-aromatic Compounds and Their Applications in Molecular Electronics: A Theoretical Investigation

Sharan Shetty,[†] Rahul Kar,[†] Dilip G. Kanhere,[‡] Sourav Pal[*†]

[†]*Theoretical Chemistry Group, Physical Chemistry Division, National Chemical Laboratory, Pune-411008, India, and* [‡]*Department of Physics, University of Pune, Pune-411007, India*

**Abstract:**

Local reactivity descriptors such as the condensed local softness and Fukui function have been employed to investigate the inter-cluster reactivity of the metallo-aromatic ($Al_4Li^-$ and $Al_4Na^-$) and anti-aromatic ($Al_4Li_4$ and $Al_4Na_4$) compounds. We use the concept of group softness and group Fukui function to study the strength of the nucleophilicity of the $Al_4$ unit in these compounds. Our analysis shows that the trend of nucleophilicity of the $Al_4$ unit in the above clusters is as follows;

$$Al_4Li^- > Al_4Na^- > Al_4Li_4 > Al_4Na_4$$

For the first time we have used the reactivity descriptors to show that these clusters can act as electron donating systems and thus can be used as a molecular cathode.

1. Introduction:

Recently, aromaticity and anti-aromaticity were shown to exist in metal clusters[1-13], thus opening a new area of metallo-aromatic and anti-aromatic compounds. Li *et al.*[1] for the first time synthesized the metallo-aromatic compounds *viz*. $Al_4Li^-$, $Al_4Na^-$, $Al_4Li_2$ etc. On the basic criteria of aromaticity, such as planarity, cyclic conjugation, Huckel's (4n+2) π electron rule, these compounds were shown to have aromaticity. Motivated with this work, Shetty *et al.*[9] predicted the possibility of anti-aromaticity in $Al_4Li_4$ metal cluster using *ab initio* calculations. At same time, in an interesting work, combining both experimental and theoretical studies, Kuznetsov *et al*[10] confirmed the existence of anti-aromaticity in $Al_4Li_3^-$ and $Al_4Li_4$ clusters. As an extension of this work, more metallo-antiaromatic compounds were discussed.[13] Very recently, magnetic properties of these compounds were investigated. Surprisingly, the magnetic properties in these compounds show somewhat different trend than the organic aromatic and anti-aromatic compounds.[14-16] The metallo-aromatic compounds were shown to have a diamagnetic ring current arising from the σ-electron delocalization rather than the π-electron, which is very negligible.[14] Thus the metal clusters are said to have σ aromaticity rather than the π aromaticity. On the other hand, metallo-anti-aromatic compounds were shown to have a diamagnetic current in the plane due to the σ electrons and a paramagnetic current out of the plane due to σ electrons.[15,16] This shows that, only on the basis of electron counting rule, it is difficult to decide the substantial existence of metallo-aromatic and anti-aromatic compounds.[16] However, this controversy is still to be resolved.[17]

The study of these kind of mixed clusters have been important in understanding the electronic and stoichiometric properties in the alloy formation. The Al-Li alloys form very strong material and are used for aerospace applications.[18] More interestingly, in the last decade, Al-Li alloys have been widely used as electron injecting material in organic light emitting diode (OLED).[19-21] It has been shown that the efficiency of the OLED's depend on the number density of electrons and holes in the organic emission layers. This can be achieved by using high work function material such as, indium tin oxide (ITO) for hole injection and Li as low work function material for electron injection.[20] Haskal *et al.* showed that the use of Li causes the organic layer to degrade due to its diffusion in the organic layer.[19] To avoid this, Al-Li alloys were used as electron injecting devices for OLED. They further showed that there is charge transfer from Li to Al at the Li-Al interface.[19] Thus, creating an ionic bond between Li and Al. This kind of bonding arrangement allows the Li-Al atoms to cluster at the interface.[19,21] Tris(8-hydroxyquinolinato) aluminum ($Alq_3$) has been the prototypical electroluminescent molecule for OLED.[21] Curioni and Andreoni[21] studied the interaction of the OLED with the metal surface such as Li, Al and Ca using *ab initio* simulations. Their investigation reveals that the $Alq_3$ accepts electron from the metal and thus acts as an electrophile. Hence, it is clear that the metal which donates electron acts as a nucleophile. The oxygen atoms in the $Alq_3$ were shown to be the electrophilic centers.[21] In an experimental work, Tang *et al* with the help of photoelectron spectroscopy, studied the charge transfer from various metal surfaces *viz.* K, Na, Ca, Mg, Al etc. They showed that the metal electronegativity is linearly dependent on the charge injection barrier.[22]

From the work of Haskal *et al.*[19] it is clear that, in the Al-Li cathode, Al acts as a nucleophile to inject electrons in the organic layer of OLED. Hence, more the nucleophilicity of Al, better its ability to donate electron to the organic layer. Motivated with this, in the present work, we use the approach of local reactivity descriptors,[23] to investigate the inter-cluster reactivity of the metallo-aromatic and anti-aromatic compounds. Furthermore, we study the strength of the nucleophilicity of the $Al_4$ unit in these compounds.

Local reactivity descriptors, such as local softness and Fukui Function (FF), based on density functional theory (DFT), have been used to determine the site reactivity in a system.[23] These descriptors have been well established for the study of intra-molecular reactivity.[24] However, it has been shown that there is some ambiguity in using these descriptors for understanding the inter-molecular reactivity for some carbonyl compounds.[24(b)] De Proft *et al*[25] have used the concept of group hardness, softness and electronegativity to study a series of organic compounds. Recently, Krishnamurty and Pal[26] used the concept of group softness for studying the inter-molecular reactivity in carbonyl compounds and organic acids. The group softness is defined as the summation of the local softness of a group of atoms around the most reactive site in a molecule.[26] It was proved from these studies that the molecule with the maximum group softness is the most reactive molecule within a series.[25, 26]

For the first time, we introduce the concept of group FF for comparing the reactivity of the metallo-aromatic and anti-aromatic compounds. The group FF is somewhat similar to the group softness defined earlier. However, group FF is defined as the group softness divided by the total softness. In this paper we study the use of both group softness and group FF for intermolecular reactivity. Furthermore, we also apply these two concepts for the first time for applications to these compounds in the field of molecular electronics. In the case of the aromatic and antiaromatic mixed metal clusters, the $Al_4$ unit acts as a superatom and hence as we will see later, the concept of group softness and group FF are extremely important.

## 2. THEORETICAL BACKGROUND:

The ground state energy of an atom or a molecule, in density functional theory, can be expressed in terms of electron density, $\rho(r)$ as [27]

$$E[\rho] = F_{HK}[\rho] + \int v(r)\rho(r)dr \qquad (1)$$

where, $v(r)$ is the external potential and $F_{HK}[\rho]$ is universal Hohenberg-Kohn functional expressed as

$$F_{HK}[\rho] = T[\rho] + V_{ee}[\rho] \qquad (2)$$

is the sum of electronic kinetic energy ($T[\rho]$) and electron-electron interaction energy ($V_{ee}[\rho]$)

The first and second partial derivative of E [$\rho$] with respect to the number of electron N under constant external potential $v(r)$ are defined as chemical potential [28], $\mu$ and the hardness [29], $\eta$ for a system.

$$\mu = \left(\frac{\partial E[\rho]}{\partial N}\right)_{v(r)} \qquad (3)$$

$$\eta = \frac{1}{2}\left(\frac{\partial^2 E[\rho]}{\partial N^2}\right)_{v(r)} = \frac{1}{2}\left(\frac{\partial \mu}{\partial N}\right)_{v(r)} \qquad (4)$$

Global softness, S defined as the inverse of hardness, can be written as

$$S = \frac{1}{2\eta} = \left(\frac{\partial N}{\partial \mu}\right)_{v(r)} \qquad (5)$$

It has been customary to use the finite difference approximation to compute $\mu$ and $\eta$ as [30]

$$\mu \approx \frac{-I-A}{2} \qquad (6)$$

$$\eta \approx \frac{I-A}{2} \qquad (7)$$

where, I and A are ionization potential and electron affinity of a chemical species, respectively.

The principle of maximum hardness was proposed by Pearson [31] relating the hardness and stability of a system at constant chemical potential and later proved by Parr and Chattaraj.[32] The global hardness reflects the overall stability of a system. However, the site selectivity and reactivity can only be studied using the local reactivity descriptors, such as local softness [23b] s(r) defined as

$$s(r) = \left(\frac{\partial \rho(r)}{\partial \mu}\right)_{v(r)} = \left(\frac{\partial \rho(r)}{\partial N}\right)_{v(r)} \left(\frac{\partial N}{\partial \mu}\right)_{v(r)} = f(r)S \qquad (8)$$

and

$$\int s(r)dr = S \qquad (9)$$

where, f(r) is the Fukui function,[23a]

$$f(r) = \left(\frac{\partial \rho(r)}{\partial N}\right)_{v(r)} = \left(\frac{\delta \mu}{\delta v(r)}\right)_N \qquad (10)$$

Thus, Fukui function can be interpreted either as the change of electron density at each point r when the total number of electron is changed or as the sensitivity of chemical potential of a system to an external perturbation at a particular point r.

Due to N discontinuity problem of atoms and molecules [33] in equation(10) leads to the introduction[23a] of both right and left hand side derivatives at a given number of electrons, $N_o(=N)$

$$f^+(r) = \left(\frac{\partial \rho(r)}{\partial N}\right)^+_{v(r)} \quad \text{for nucleophilic attack, and} \quad (11a)$$

$$f^-(r) = \left(\frac{\partial \rho(r)}{\partial N}\right)^-_{v(r)} \quad \text{for electrophilic attack} \quad (11b)$$

By finite difference method using the electron densities of No, No+1, No-1 electron systems, Fukui functions for the nucleophilic and electrophilic attack can be defined respectively, as

$$f^+(r) \approx \rho_{N_o+1}(r) - \rho_{N_o}(r) \quad (12a)$$

$$f^-(r) \approx \rho_{N_o}(r) - \rho_{N_o-1}(r) \quad (12b)$$

and for radical attack,

$$f^o(r) \approx \frac{1}{2}\left(\rho_{N_o+1}(r) - \rho_{N_o-1}(r)\right) \quad (12c)$$

In order to describe the site reactivity or site selectivity, Yang *et al.*[34] proposed atom condensed Fukui function, based on the idea of electronic population around an atom in a molecule, similar to the procedure followed in population analysis technique.[35] The condensed Fukui function for an atom k undergoing nucleophilic, electrophilic or radical attack can be defined respectively as

$$f_k^+ \approx q_k^{N_o+1} - q_k^{N_o} \quad (13a)$$
$$f_k^- \approx q_k^{N_o} - q_k^{N_o-1} \quad (13b)$$
$$f_k^o \approx \frac{1}{2}\left(q_k^{N_o+1} - q_k^{N_o-1}\right) \quad (13c)$$

where $q_k$'s are electronic population of the kth atom of a particular species.
The condensed local softness, $s_k^+$ and $s_k^-$ are defined accordingly for nucleophilic and electrophilic attack, respectively.

In many cases, the interaction takes place not just with a single atom, but a group of atoms. In such cases group softness and group FF can be defined as a sum of the local softness or FF of the group of atoms participating in the interaction. The group softness, $s_g$ is defined over the set of the atoms $n_g$ is written as

$$s_g = \sum_{k=1}^{n_g} s_k \quad (14)$$

This concept is very useful when reaction does not take place through a single atom, but through neighbouring atoms in a cooperative manner, in other words, in cases where a group of atoms participate in bonding together. We can also define a concept of group FF, $f_g$, which is nothing but the sum of the FF of the atoms of the group. It is obvious that the $f_g$ is nothing but $\frac{s_g}{S}$, where S is the total softness of the system. For intra-molecular reactivity, however, $f_g$ and $s_g$ will contain the same information. On the other hand, in comparing reactivity across different molecules, i.e. in comparing inter-molecular, $f_g$ and $s_g$ will not necessarily give the same trend, since global softness of the systems vary.

Along the lines of electrophilic and nucleophilic condensed FF, we can define the electrophilic and nucleophilic condensed group softness and group FF, $s_g^{+/-}$ and $f_g^{+/-}$ etc. For the systems of metal clusters that are being studied, $Al_4$ unit of atoms behave as a superatom in acceptance of electrons and hence the concept of group softness or group FF will be extremely useful.

3. **Computational Details**:

Geometry optimization of the metallo-aromatic and anti-aromatic clusters *viz.* $Al_4Li^-$, $Al_4Na^-$, $Al_4Li_4$ and $Al_4Na_4$ were performed at the MP2 level using 6-31 +G* basis set. The structural parameters of these clusters are in good agreement with the earlier work, which is discussed in the next section. Chemical potential, global hardness and global softness are calculated from equations 5, 6 and 7. The local reactivity parameters such as condensed local softness and condensed FF are obtained from the above equations using Lowdin population analysis (LPA)[36]. All the calculations in the present work have been performed by using GAMESS software.[37]

4. Results and Discussion:

We begin this section with a brief discussion of the structural properties of the metallo-aromatic and anti-aromatic compounds *viz.* $Al_4Li^-$, $Al_4Na^-$, $Al_4Li_4$ and $Al_4Na_4$. Later, we discuss the use of group FF to analyze the reactivity in these systems.

## I. Structure

The ground state geometries of $Al_4Na^-$, $Al_4Li_4$ and $Al_4Na_4$ are shown in Fig. 1. The ground state geometry of $Al_4Li^-$ has also a pyramidal structure similar to $Al_4Na^-$ as shown in the Fig. 1. For more detailed study on the stability of these clusters, studies by Shetty *et al*[13(a)] and Kuznetsov *et al*[10] may be referred. The ground state of $Al_4Li^-$ and $Al_4Na^-$ metallo-aromatic clusters shows that the $Al_4$ unit has a square structure with the Al-Al bond lengths being 2.61 and 2.62 A. However, the $Al_4$ unit in the $Al_4Li_4$ and $Al_4Na_4$ metallo-anti-aromatic clusters has a rectangular geometry (Fig 1 (b) and (c)). The bond lengths of these clusters obtained from MP2 calculations in the present work are in good agreement with the earlier results.[1,13]

It is worth mentioning that in all the metallo-aromatic and anti-aromatic compounds, the Al atoms group to form one $Al_4$ unit in the whole cluster. This arrangement allows the $Al_4$ unit to form a superatom.[38] Interestingly, Shetty *et al* have discussed the role of $Al_4$ superatom in the $Al_4Na_4$ and $Al_4Na_3^-$ metallo-antiaromatic compounds.[13] It should be noted that if there is no formation of $Al_4$ superatom in the metallo-aromatic and anti-aromatic compounds, such as $Al_4Li_2$, $Al_4Li_4$, $Al_4Na_4$ etc., the charge transfer from the Li and Na atoms to the $Al_4$ unit would be difficult. Hence, the existence of aromaticity and anti-aromaticity in these compounds is not possible. A careful study of charge transfer and optical properties in the metallo-anti-aromatic compounds have been discussed by Datta *et al*.[39]

## II. Reactivity

We now focus our discussion on the inter-cluster reactivity of the metallo-aromatic and anti-aromatic compounds. We investigate the trend of the nucleophilicity of the $Al_4$ unit in all the four clusters *viz*. $Al_4Li^-$, $Al_4Na^-$, $Al_4Li_4$ and $Al_4Na_4$ using the group softness and FF.

### a. Metallo-aromatic compounds: $Al_4Li^-$, $Al_4Na^-$

Table 1 gives the chemical potential, global hardness and global softness values of $Al_4Li^-$, $Al_4Na^-$, $Al_4Li_4$ and $Al_4Na_4$ clusters. The local reactivity parameters such as the condensed local softness for the electrophilic ($s^-$) and the nucleophilic ($s^+$) attack, condensed FF for the electrophilic ($f^+$) and the nucleophilic ($f^-$) attack , group softness ($s^-$) and the group FF ($f^-$) for the electrophilic attack of the Al$_4$ unit in the Al$_4$Li$^-$, Al$_4$Na$^-$ clusters are given in Table 2 and 3 respectively. Since, we are only interested in the strength of the nucleophilicity (for the electrophilic attack) of the Al$_4$ unit, we only discuss $s_g^-$ and $f_g^-$ for the electrophilic attack on the Al$_4$ unit i.e. the nucleophilicity of the Al$_4$ unit.

It is seen from the Table 2 that the condensed local softness and the condensed FF values for the electrophilic ($f^-$, $s^-$) and the nucleophilic attack ($f^+$, $s^+$) of all the 4 Al atoms of the Al$_4$Li$^-$ are identical. This is due to the electron delocalization in the Al$_4$ unit of Al$_4$Li$^-$ system. Hence, all the 4 Al atoms have the same reactivity. The Al$_4$ species in the Al$_4$Na$^-$ cluster also has a similar behavior to the Al$_4$Li$^-$ cluster.

We can clearly see from Table 2 and 3 that the Al$_4$ unit of the Al$_4$Li$^-$ has higher value of $s_g^-$ than the Al$_4$Na$^-$. More interestingly, the trend of the group FF for these systems is also similar (Table 2and 3). Hence, the Al$_4$ unit of the Al$_4$Li$^-$ cluster can act as a better electron donating system than the Al$_4$Na$^-$ cluster.

**b. Metallo-anti-aromatic compounds: Al$_4$Li$_4$ and Al$_4$Na$_4$**

The local reactivity parameters of the Al$_4$Li$_4$ and Al$_4$Na$_4$ clusters are given in Table 4 and 5 respectively. The condensed local softness and FF of the metallo-anti-aromatic compounds show a different trend than the metallo-aromatic compounds.

From Table 4 and 5, we see that the values of the local softness and the FF for the electrophilic attack of all the four Al atoms of the Al$_4$Li$_4$ and Al$_4$Na$_4$ clusters are same. Surprisingly, the values of the condensed local softness and condensed FF for the nucleophilic attack of two of the Al atoms are different than the other two Al atoms (Table 4 and 5). This shows that the ability to donate the electrons of all the four Al atoms in both the clusters is same. However, the ability to accept the electrons of two Al atoms is different than the other two. The reason for this may be the electron localization along the shorter Al-Al bonds.

The group softness and FF results show that, even in the metallo-anti-aromatic compounds the $Al_4$ unit of the $Al_4Li_4$ cluster has higher group softness and FF than the $Al_4Na_4$ cluster.

If we compare the strength of nucleophilicity of the $Al_4$ unit of all the four clusters *viz*. $Al_4Li^-$, $Al_4Na^-$, $Al_4Li_4$ and $Al_4Na_4$ clusters, the trend is as follows.
$Al_4Li^- > Al_4Na^- > Al_4Li_4 > Al_4Na_4$
The above discussion shows that the nucleophilicity of the Al atoms decides the work function of the Al-Li cathode. Hence, we can say that, higher is the nucleophilicity of the Al atoms, lower is the work function of the Al-Li cathode.

Curioni *et al.* proved in their study that the oxygen atoms of the $Alq_3$ molecule are the electron withdrawing groups (electrophiles).[21] Hence, from the present study one can consider the metallo-aromatic and anti-aromatic systems to be molecular cathodes for injecting electrons in a single electroluminescent molecule, such as, $Alq_3$.

**Conclusion**

In the present work for the first time, we have compared the inter-cluster reactivity of metallo-aromatic and anti-aromatic compounds using the local reactivity descriptors. The results show that the $Al_4$ unit in all the four clusters *viz*. $Al_4Li^-$, $Al_4Na^-$, $Al_4Li_4$ and $Al_4Na_4$ acts as a nucleophile. The inter-cluster reactivity suggests that the highest nucleophilicity of the $Al_4$ unit is in the $Al_4Li^-$ cluster. Moreover, for the first time we show that these clusters can be probable candidates for molecular cathodes.

**Acknowledgement**

S. Shetty and S. Pal gratefully acknowledge the Indo-French Center for the Promotion of Advance Research (IFCPAR) (Project. No. 2605-2), New Delhi, India, for financial assistance. R. Kar gratefully acknowledges the Council for Scientific and Industrial Research (CSIR), New

Delhi for financial assistance.
†National Chemical Laboratory

‡University of Pune

* Author to whom correspondence may be addressed. Fax: +91-20-5893044. E-mail: s.pal@ncl.res.in.

**Table 1:** Chemical potential, global hardness and global softness of $Al_4Li^-$, $Al_4Na^-$, $Al_4Li_4$, $Al_4Na_4$.(values are in atomic units)

| Cluster | chemical potential, $\mu$ | Global hardness, $\eta$ | global Softness, S |
|---|---|---|---|
| Al4Na- | -0.0055 | 0.0695 | 7.194 |
| Al4Li- | -0.0055 | 0.0775 | 6.452 |
| Al4Na4 | -0.0565 | 0.0970 | 5.128 |
| Al4Li4 | -0.0845 | 0.1128 | 4.405 |

**Table 2**: Local reactivity parameters of $Al_4Li^-$. Condensed local softness for nucleophilic ($s^+$), electrophilic ($s^-$) attack and respective group softness $s_g^+$ and $s_g^-$. Condensed Fukui function for nucleophilic ($f^+$), electrophilic ($f^-$) attack and respective group Fukui functions $f_g^+$ and $f_g^-$.

| Atom | $s^+$ | $s^-$ | $s_g^+$ | $s_g^-$ | $f^+$ | $f^-$ | $f_g^+$ | $f_g^-$ |
|---|---|---|---|---|---|---|---|---|
| Al | 0.129 | 1.336 | | | 0.020 | 0.207 | | |
| Al | 0.129 | 1.336 | 0.516 | 5.344 | 0.020 | 0.207 | 0.080 | 0.828 |
| Al | 0.129 | 1.336 | | | 0.020 | 0.207 | | |
| Al | 0.129 | 1.336 | | | 0.020 | 0.207 | | |

**Table 3:** Local reactivity parameters of $Al_4Na^-$ cluster in atomic units. Condensed local softness for nucleophilic ($s^+$), electrophilic ($s^-$) attack and respective group softness $s_g^+$ and $s_g^-$. Condensed Fukui function for nucleophilic ($f^+$), electrophilic ($f^-$) attack and respective group Fukui functions $f_g^+$ and $f_g^-$.

| Atom | $s^+$ | $s^-$ | $s_g^+$ | $s_g^-$ | $f^+$ | $f^-$ | $f_g^+$ | $f_g^-$ |
|---|---|---|---|---|---|---|---|---|
| Al | 0.201 | 1.388 |       |       | 0.028 | 0.193 |       |       |
| Al | 0.201 | 1.388 | 0.804 | 5.552 | 0.028 | 0.193 | 0.112 | 0.772 |
| Al | 0.201 | 1.388 |       |       | 0.028 | 0.193 |       |       |
| Al | 0.201 | 1.388 |       |       | 0.028 | 0.193 |       |       |

**Table 4:** Local reactivity parameters of $Al_4Li_4$ cluster in atomic units. Condensed local softness for nucleophilic ($s^+$), electrophilic ($s^-$) attack and respective group softness $s_g^+$ and $s_g^-$. Condensed Fukui function for nucleophilic ($f^+$), electrophilic ($f^-$) attack and respective group Fukui functions $f_g^+$ and $f_g^-$.

| Atom | $s^+$ | $s^-$ | $s_g^+$ | $s_g^-$ | $f^+$ | $f^-$ | $f_g^+$ | $f_g^-$ |
|---|---|---|---|---|---|---|---|---|
| Al | 0.366 | 0.542 | | | 0.083 | 0.123 | | |
| Al | 0.366 | 0.542 | 1.420 | 2.168 | 0.083 | 0.123 | 0.322 | 0.492 |
| Al | 0.344 | 0.542 | | | 0.078 | 0.123 | | |
| Al | 0.344 | 0.542 | | | 0.078 | 0.123 | | |

**Table 5:** Local reactivity parameters of Al$_4$Na$_4$ cluster in atomic units. Condensed local softness for nucleophilic ($s^+$), electrophilic ($s^-$) attack and respective group softness $s_g^+$ and $s_g^-$. Condensed Fukui function for nucleophilic ($f^+$), electrophilic ($f^-$) attack and respective group Fukui functions $f_g^+$ and $f_g^-$.

| Atom | $s^+$ | $s^-$ | $s_g^+$ | $s_g^-$ | $f^+$ | $f^-$ | $f_g^+$ | $f_g^-$ |
|---|---|---|---|---|---|---|---|---|
| Al | 0.415 | 0.487 | | | 0.081 | 0.095 | | |
| Al | 0.133 | 0.487 | 1.096 | 1.948 | 0.026 | 0.095 | 0.214 | 0.380 |
| Al | 0.415 | 0.487 | | | 0.081 | 0.095 | | |
| Al | 0.133 | 0.487 | | | 0.026 | 0.095 | | |

**Figure Captions.**

Figure. 1 Optimized equilibrium geometries of (a) $Al_4Na^-$ (b) $Al_4Li_4$ (c) $Al_4Na_4$ at MP2 level with 6-31+G*.

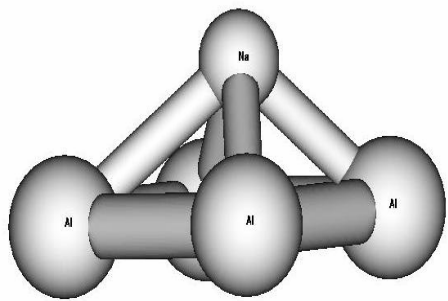

**(a)**

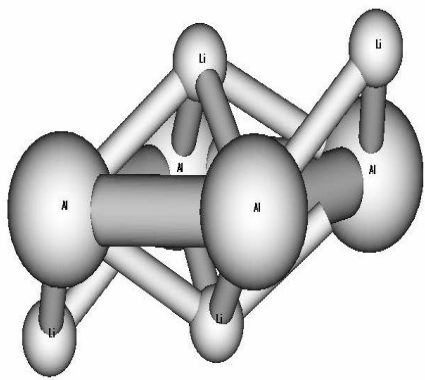

**(b)**

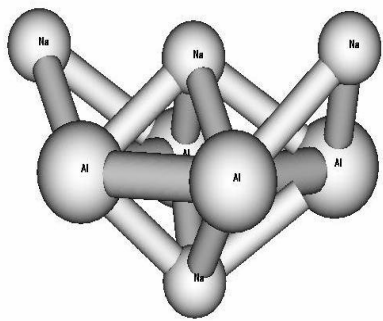

**(c)**
**Figure. 1**

## Table of contents

**Sharan Shetty, Rahul Kar, Dilip G. Kanhere,[‡] Sourav Pal[*†]**

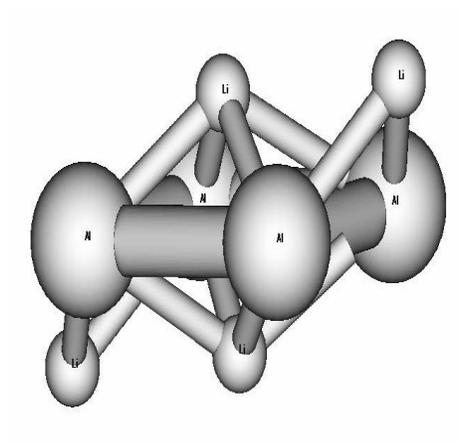